# Verification of Space Weather Forecasts issued by the Met Office Space Weather Operations Centre

**M. A. Sharpe[1], S. A. Murray[2]**

[1]Met Office, UK. [2]Trinity College Dublin, Ireland.

(michael.sharpe@metoffice.gov.uk)

**Key Points:**

- An operational real time verification system has been developed for geomagnetic storm and solar flare forecasts.
- Ranked (probability) skill score, ROC plots, and reliability diagram analyses are shown for a period between April 2015 and October 2016.
- Both geomagnetic storms and flares were generally over-forecast.
- The skillful rolling prediction periods of 180 and 120 days were identified for geomagnetic and flare forecasts, respectively.

## Abstract

The Met Office Space Weather Operations Centre was founded in 2014 and part of its remit is a daily Space Weather Technical Forecast to help the UK build resilience to space weather impacts; guidance includes four day geo-magnetic storm forecasts (GMSF) and X-ray flare forecasts (XRFF). It is crucial for forecasters, users, modelers and stakeholders to understand the strengths and weaknesses of these forecasts; therefore, it is important to verify against the most reliable truth data source available. The present study contains verification results for XRFFs using GOES-15 satellite data and GMSF using planetary K-index (Kp) values from the GFZ Helmholtz Centre. To assess the value of the verification results it is helpful to compare them against a reference forecast and the frequency of occurrence during a rolling prediction period is used for this purpose. Analysis of the rolling 12-month performance over a 19-month period suggests that both the XRFF and GMSF struggle to provide a better prediction than the reference. However, a relative operating characteristic and reliability analysis of the full 19-month period reveals that although the GMSF and XRFF possess discriminatory skill, events tend to be over-forecast.

## 1 Introduction

In recent decades there have been significant technological advances upon which governments, industries and organizations have become increasingly dependent. Many of these advances are vulnerable to space weather to the extent that security and/or safety could be severely compromised when significant events occur. After severe space weather was added to the UK's National Risk Register of Civil Emergencies in 2011, the UK government sought to establish a 24/7 space weather forecasting centre and the Met Office Space Weather Operations Centre (MOSWOC) was officially opened on 8$^{th}$ October 2014. Part of MOSWOCs remit is to issue a daily Space Weather Technical Forecast (SWTF) to help affected UK industries and infrastructure build resilience to space weather events; issued at midnight with a midday update, it contains a:

- space weather activity analysis;
- four-day solar activity summary;
- geo-magnetic storm forecast (GMSF);
- coronal mass ejection (CME) warning service;
- X-ray flare forecast (XRFF);
- solar radiation storm forecast; and
- high energy electron event forecast.

Verification of these products is crucially important for forecasters, users, modelers and stakeholders because it facilitates an understanding of the strengths and weaknesses of each forecast product. Ideally verification should be performed in near-real time to enable instant forecaster feedback because this enables:

a) necessary corrections to be made in a timely fashion; and
b) operational forecasters to use the results to further develop their forecasting skills.

As a member of the International Space Environment Service (http://www.spaceweather.org), the Met Office is helping to coordinate verification efforts with

1. NASA Community Coordinated Modelling Center on the implementation of the Flare Scoreboard (https://ccmc.gsfc.nasa.gov/challenges/flare.php); a system to enable the automatic upload of flare predictions and provide immediate verification to inter-compare the forecasts from participating organisations;
2. the EU on the development project Flare Likelihood And Region Eruption foreCASTing (FLARECAST; http://www.flarecast.eu) to automatically forecast and verify X-ray flares.

Some initial MOSWOC flare forecast verification has been undertaken as part of the FLARECAST project (*Murray et al*, 2017), however the real-time operational verification system that has now been developed for use by the MOSWOC forecasters was not fully explored in this work. Operational verification of most SWTF products is planned and progress to date includes investigations into the skill of GMSFs, XRFFs and Earthbound-CME warnings; with near-real-time verification of the former two products already operational using the Warnings Verification System (WVS) (*Sharpe*, 2015) and the Area Forecast Verification System (AFVS) (*Sharpe*, 2013). The methodologies used to verify GMSFs and XRFFs are outlined in Section 2 and results for the period between April 2015 and October 2016 are presented in Section 3. Section 4 contains brief conclusions and an outline of further work.

## 2 Verification Methodologies

2.1 Geo-Magnetic Storm Forecasts

The GMSF is both probabilistic and multi-category; each category referring to a different geo-magnetic activity level, measured using the K index at 13 observing sites stationed across the globe, from which a planetary K value (Kp) is evaluated. GMS level:

> G1/G2. (Minor/Moderate) denote Kp values of 5-, 5o, 5+, 6-, 6o or 6+;
>
> G3. (Strong) denote Kp values of 7-, 7o or 7+;
>
> G4. (Severe) denote Kp values of 8-, 8o or 8+; and
>
> G5. (Extreme) denote Kp values of 9-, 9o or 9+.

Additionally G0, denotes Kp values of 4+ or below.

Forecasters issue GMSFs by first analyzing images to identify CMEs and coronal holes and then using the Wang-Sheeley-Arge Enlil model (*Edmonds, 2013*) to identify high-speed solar wind streams and CMEs. However associated forecasts of GMSs are limited because values of the z–component of the sun's magnetic field are unknown (except as measured by the ACE/DSCOVR satellite). One further source of information is from an Autoregressive Integrated Moving Average model of Kp values; however, there is no model which accurately predicts Kp fluctuations. Consequently, forecasting is essentially a subjective process which continues to rely heavily upon the experience of operational forecasters.

Table 1 shows the GMSF within the 00Z issue of the SWTF on 1st October 2016.

| GMS Probability (Exceedance) | Level | Past 24 Hours (Y/N) | Day 1 (00-24 UTC) (%) | Day 2 (00-24 UTC) (%) | Day 3 (00-24 UTC) (%) | Day 4 (00-24 UTC) (%) |
|---|---|---|---|---|---|---|
| Minor or Moderate | G1 to G2 | Y | 55 | 35 | 30 | 10 |
| Strong | G3 | N | 5 | 5 | 1 | 1 |
| Severe | G4 | N | 1 | 1 | 1 | 1 |
| Extreme | G5 | N | 1 | 1 | 1 | 1 |

**Table 1.** GMSF contained within the 00Z SWTF issued on 1st October 2016.

The columns from left to right display:

1. a single word description of the GMS type;

2. the G-scale level associated with this type of GMS;

3. whether each type of GMS has been observed during the previous 24 h period;

4-7. forecast probabilities for the likelihood that each type of GMS will occur during four consecutive days into the future.

During day 1 (Table 1 column 4) the predicted probability of a GMS level: ≥ G1 was 55%, ≥G3 was 5%, ≥ G4 was 1% and ≥ G5 was 1%. However, the minimum forecast probability is stipulated at 1%; therefore in this analysis a forecast value of 1% is interpreted as 0%.

The probabilities in Table 1 refer to the chance that the GMS level will be reached or exceeded at least once during the 24 h period. Therefore, column 4 forecasts that the probability associated with:

G0. is 100% - 55% = 45%;

G1/G2. is 55% - 5% = 50%;

G3. is 5% - 0% = 5%;

G4. and G5. is 0%.

GFZ Helmholtz Centre Potsdam Kp values are used as the truth data source for GMSF verification; however, these values have a one-month latency period which makes them unsuitable for near-real-time verification. Consequently, estimated Kp values from the Space Weather Prediction Center (SWPC) are used to provide valuable feedback to MOSWOC forecasters. During the first ten months of 2016, SWPC and GFZ Kp values differed on 81 days.

The verification metric most commonly associated with multi-category probabilistic forecasts is the Ranked Probability Score (*RPS*) (*Epstein*, 1969 and *Murphy*, 1971). The RPS is defined by

$$RPS = \sum_{n=0}^{5}(P(G_n) - O(G_n))^2; \tag{1}$$

where in the present case $P(G_n)$ is the forecast probability that the maximum GMS level to be observed during the 24h period is $\leq G_n$ (where $n = 0, 1/2, 3, 4$ or $5$) and $O(G_n)$ is 0 if the maximum observed level is $< G_n$ and 1 otherwise. The $RPS$ (which ranges from 1 to a perfect score of 0) is calculated separately for every day of each forecast and a mean value ($\overline{RPS}$) is obtained by simply averaging the $RPS$ values calculated for a large number of forecasts; 90% confidence intervals are produced using simple bootstrapping with replacement. The $RPS$ provides a very valuable approach to the problem of verifying multi-category probabilistic forecasts; however, a reference is required against which to benchmark the performance. Three common reference forecast choices are: random chance, persistence and climatology. Short-term climatology (subsequently referred to as a prediction-period) has been chosen for the reference in the present study and $RPS_{ref}$ has been evaluated by replacing $P(G_n)$ in Equation (1) with the frequency of occurrence of GMSs over a prediction period encompassing the most recent 180 days. 180 days was chosen following an investigation (outlined in Section 3) which revealed it to be an accurate predictor period for GMSs. The use of a reference forecast enables the Ranked Probability Skill Score ($RPSS$), defined by

$$RPSS = 1 - \frac{\overline{RPS}}{\overline{RPS}_{ref}} \qquad (2)$$

to be evaluated; this score ranges from -∞ to a perfect score of 1 with $RPSS > 0$ implying that the forecast is more skilful than the reference. Confidence intervals for this statistic (calculated using bootstrapping with replacement) indicate whether there is any statistically significant evidence to suggest that the forecast is more skillful than the reference.

Verification of the GMSF has been performed for each forecast range by the AFVS (*Sharpe*, 2013) using daily maximum values of Kp. The AFVS was originally designed to verify a range of forecast categories against a truth data distribution (representing the conditions throughout an area); however, when presented with a truth data source containing only a daily maximum it can also be used to verify a daily maximum forecast like the GMSF.

An alternative verification approach is to treat the GMSF as a probabilistic warning service, verifying the forecast probabilities associated with each GMS level separately. In practice however, only categories ≥ G1 may be evaluated, because the more severe levels occur so rarely that robust statistics cannot be obtained over the available time frame. In the present study the Warnings Verification System (WVS) (*Sharpe*, 2015) has been used to verify the GMSF as a service, using Relative Operating Characteristic (ROC) plots and reliability diagrams (*Jolliffe and Stephenson*, 2012). The WVS is a flexible system originally developed to verify terrestrial weather; this system allows the analysis of near-hits by way of flexing thresholds in terms of space, time, intensity and confidence. However, in the present study flexing has only been applied in terms of intensity so that Kp values of 4-, 4o and 4+ are each categorized as a 'low-miss' except when they occur during a warning when they are categorized as a 'low-hit'. The only other possible flex appropriate to this analysis is time because: confidence flexing cannot be applied since there is only one definitive Kp value and spatial flexing cannot be applied since no near-Earth Kp values are available against which to assess the forecast.

| X Ray Flares Probability | Level | Past 24 Hours (Y/N) | Day 1 (00-24 UTC) (%) | Day 2 (00-24 UTC) (%) | Day 3 (00-24 UTC) (%) | Day 4 (00-24 UTC) (%) |
|---|---|---|---|---|---|---|
| Active | M Class | Y | 28 | 28 | 28 | 28 |
| Very Active | X Class | N | 2 | 2 | 2 | 2 |

**Table 2.** XRFF contained within the 00Z SWTF on 21st July 2016.

2.2 X-ray Flare Forecasts

The second aspect of the SWTF considered in the present study is the X-ray Flare Forecast (XRFF); a sample of which is shown in Table 2.

This forecast is similarly displayed to the GMSF shown in Table 1; from left to right, column(s):

1. contains a description of the type of XRF;

2. identifies the class associated with each type of flare;

3. identifies whether any XRFs have been observed during the previous 24 h period;

4-7. contain forecast probabilities that each type of XRF will occur during four consecutive days into the future.

Forecast probabilities for each active region are calculated using a Poisson-statistics technique (*Gallagher et al, 2002*) based on historical flare rates for forecaster-assigned McIntosh classifications (*McIntosh, 1990*). These active region probabilities are combined to give a full-disk forecast, i.e., the chance of a flare occurring somewhere on the solar disk in the next 24 hours. The resulting model probability can be edited by the MOSWOC forecaster using their expertise before being issued as the Day 1 forecast as shown in Table 2. The Day 2-4 forecasts are purely based on forecaster expertise. More details about the forecasting method can be found in Murray et al, 2017. The XRF classes below M-class are A-class, B-class and C-class; however, these types of flare are not included in the forecast. In the soft X-ray range, flares are classified as A-, B-, C-, M-, or X- class according to the peak flux measured near Earth by the GOES spacecraft over 1-8 Å (in $Wm^{-2}$). Each class has a peak flux ten times greater than the preceding one, with X-class flares having a peak flux of order $10^{-4}$ $Wm^{-2}$." During each 24 h period of Table 2 an M-class flare is predicted to occur with a probability of 28% and an X-class flare with a probability of 2%.

There is a subtle, yet important difference between the values contained within Table 1 and Table 2; in the former the probabilities denote the chance of exceeding each level, whereas in the latter the probabilities indicate the chance that each class will be observed at least once during the 24 h period. Consequently, using $P(X)$ and $P(M)$ to denote the probabilities associated with X-class and M-class flares (as they appear in Table 2), it is theoretically possible for $P(X)$ to be greater than $P(M)$; whereas, $P(G5) > P(G4)$ is impossible. Although the values in Table 2 denote the probabilities of observation (rather than exceedance), user interest will lie mainly in the maximum flux class to occur during each 24 h period; therefore, some manipulation of the values displayed in Table 2 is required. The following paragraph derives expressions for

$P$(maximum flux is A,B or C-class), $P$(maximum flux is M-class) and
$P$(maximum flux is X-class) from the probabilities that appear in Table 2 (denoted by $P(M)$ and $P(X)$).

Evaluating the probability that the maximum flux is X-class is simple; because X is the maximum possible flux class;

$$P(\text{maximum flux is X-class}) = P(X). \quad (1)$$

To calculate the probability that the maximum flux is M-class it is first necessary to observe that

$$P(\bar{M}) = P(\text{minimum flux is X-class}) + P(\text{maximum flux is A, B or C-class}) \quad (2)$$

where $P(\bar{M}) = 1 - P(M)$ is the probability that M-class will not occur. The XRF truth data source is long wave radiation observations reported by the Geo-Orbiting Earth Satellite (GOES-15) which takes measurements every 60-seconds. X-class flares are very rare, for example Wheatland et al (2005) noted that out of 10,226 days from 1975 to 2003, M-class flares occurred on ~25% of those days whereas X-class events occurred on only ~4% of those days. Analysis for the present study reveals that X-class flares occurred on just over 2% of days between 2010 and 2015. Therefore, it is relatively safe to assume that the first term on the right hand side of Equation (2) is zero since it is virtually impossible for the minimum 60-second observation during a 24 h period to be X-class and consequently,

$$P(\text{maximum flux is A, B or C-class}) = 1 - P(M). \quad (3)$$

The final expression to obtain is $P$(maximum flux is M-class); since

$$P(\text{maximum flux is M-class}) = 1 - P(\text{maximum flux is not M-class})$$

it follows that

$$P(\text{maximum flux is M-class}) = 1 - (P(X) + P(\text{Maximum flux is A, B or C-class}))$$

which, on application of Equation (3), gives

$$P(\text{maximum flux is M-class}) = P(M) - P(X). \quad (4)$$

Equations (1), (3) and (4) are used to calculate maximum XRFF probabilities which are used by the AFVS to verify the skill using the RPS and the RPSS. For the latter, a reference forecast is necessary and (as with the GMSF) the frequency of XRF class activity during a rolling prediction period is used for this purpose. The analysis outlined in Section 3 suggest that a prediction period of 120 days is a suitable choice. The XRFF is also assessed by the WVS, verifying separately the forecast probabilities associated with M-class and X-class flares; in practice however, only M-class flares can be evaluated because during the trial period X-class flares occurred too rarely to facilitate the calculation of robust statistics. As was the case for GMS verification, only intensity flexing is applied, for which a low-hit threshold of $10^{-6} Wm^{-2}$ (C-class) is used.

# 3 Results

Two 4-day SWTFs are issued each day; a main forecast is issued at 00Z and an update at 12Z. It is inappropriate to analyze the performance on day 1 by amalgamating the update with the main forecast because the day 1 component to this update is a 12 h (rather than a 24 h) forecast; therefore, only 00Z forecasts are considered in the present study.

3.1 Geo-Magnetic Storms

The RPS is used to assess the skill of MOSWOC forecasts; however, as discussed in Section 2, a reference forecast is required against which to benchmark the score by calculating the RPSS. Arguably the simplest (and most basic) choice is random chance since its production requires no prior knowledge or information. The skill associated with a forecast generated by random chance is usually low and consequently it does not predict events well. However, despite this random chance is used (implicitly) in a number of popular verification statistics such as the Equitable Threat Score and the Peirce and Heidke Skill Scores (*Jolliffe and Stephenson*, 2012). Persistence is another popular reference choice, again (no doubt) because it requires little prior knowledge or information; persistence forecasts usually predict that tomorrow will be the same as today. When events are rare or conditions are benign persistence can produce a very favorable score; however, it is a completely ineffective predictor of the onset of severe events. The third most popular choice for a reference forecast is climatology. This option is less common because it requires prior knowledge of the conditions over a long time frame; indeed, in meteorology it is common to calculate climatology over a 30 years period. This reference sets the probability of each forecast category to its climatological frequency of occurrence. Usually the climatological period is fixed in advance; for example, in meteorology it is common to compare the latest season against the distribution formed by accumulating each corresponding season over the 30-year period between 1981 and 2010 (*National Climate Information Centre*, 2017). The climatology of solar activity is dissimilar to meteorological climatology, so it is not valid to follow this methodology exactly; however, it is unclear which prediction period is most appropriate. Consequently, different period lengths (from 30 to 360 days) have been analyzed to obtain an accurate predictor. For each prediction period definitive GFZ data was used to calculate the frequency of occurrence of each GMS-level; however, GFZ Kp values are only available following a one-month (approximately) latency period (although SWPC produce real time estimates). The near-real time nature of GMS forecast verification precludes the use of truth data which is unavailable to the forecaster; therefore, a one-month latency period has been built into this analysis. Extensive checking confirmed that each prediction period length appeared to produce a similar 12-month rolling mean RPS value. Therefore, the minimum RPS value was calculated on only the first day of every month throughout the 16-year trial period. This method identified 180 days as the best performing prediction period length . Figure 1 (which uses definitive Kp-values from GFZ) displays the rolling 180-day frequency of occurrence of G0 (×), G1-2 (×), G3 (□) and G4 (□), and G5 (o) for this 16-year trial period (beginning January in 2001) which also includes the period of MOSWOC forecasts analyzed in the present study.

Figure 1 clearly reveals the extent to which G0 dominates, although there has been a decrease in its frequency in recent months; in the 180 day period to February 2016 G0 was the maximum GMS level to occur on only 78.4% of occasions, down from a previous high of 96.0% in the 180 day period to October 2014. During the period of SWTF analysis (April 2015 to October 2016) the maximum GMS level to occur in the 180 day prediction period was:

- G0 on between 77.8% and 88.9% of days;
- G1-2 on between 11.1% and 19.4% of days;

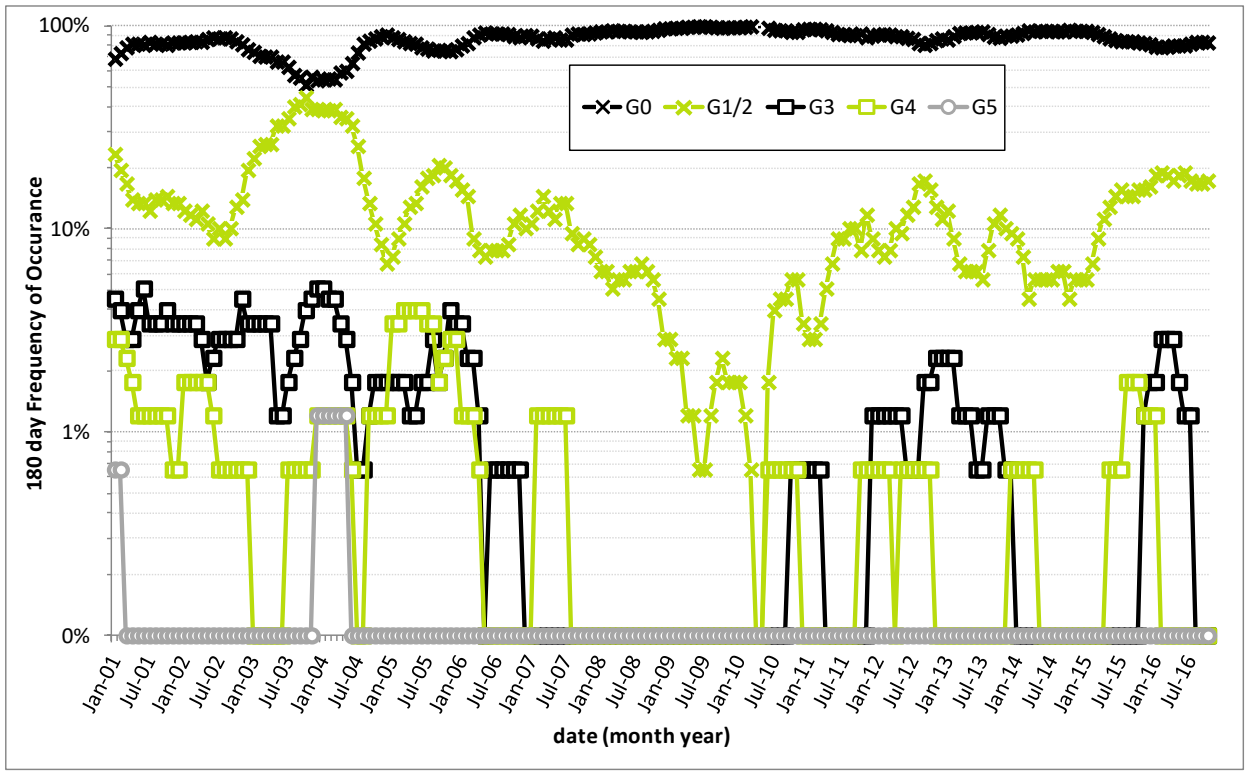

**Figure 1.** Rolling 180-day frequency of occurrence of daily maximum GMS level.

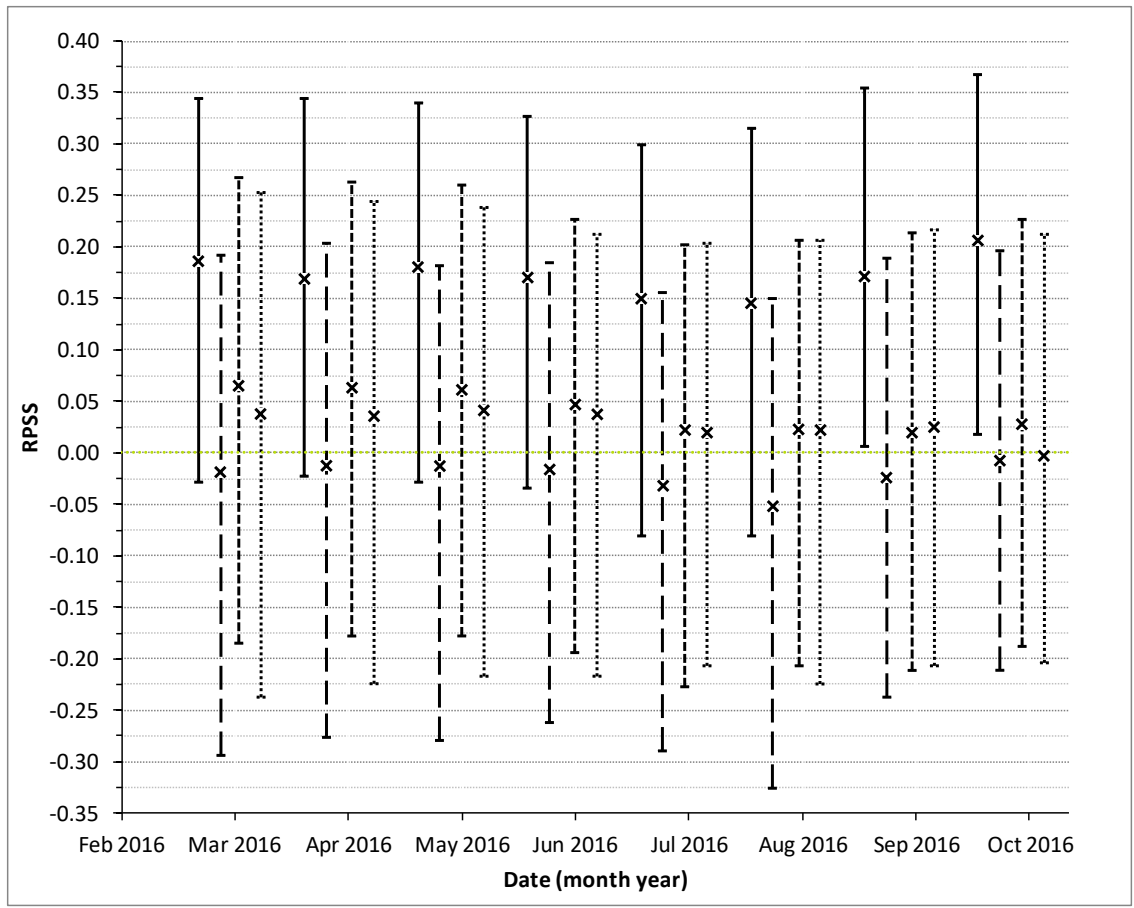

**Figure 2.** Rolling 12-monthly RPSS values (×) with 90% bootstrapped confidence intervals for each day of the GMSF. Days 1 to 4 are indicated by solid, long dashed, short dashed and dotted lines, respectively.

- G3 on between 0% and 2.8% of days;
- G4 on between 0% to 1.7% of days; and
- G5 was not observed.

Although it is inappropriate to calculate the RPSS for individual forecasts, monthly or annual values are available via Equation (2) following an evaluation of $\overline{RPS}$ and $\overline{RPS}_{ref}$ - the latter being calculated by substitution of the PDFs in Figure 1 for the forecast.

Figure 2 displays rolling 12-monthly RPSS estimates (×) and 90% confidence intervals for each day of the GMSF throughout the April 2015 to October 2016 period of analysis. Each point in this figure is plotted against the final month of the 12-month period to which it corresponds, consequently March 2016 represents the period from April 2015 to March 2016, and October 2016 represents the period from November 2015 to October 2016. Although the RPSS is greatest on day 1 the majority of the confidence intervals associated with it cross the green no-skill line. The lower tail of the intervals for October 2016 does not intersect with this line, implying evidence at the 90% level to indicate that the skill during this 12-month period is greater than that associated with a rolling 180-day prediction period of GMS activity. However, all remaining confidence intervals intersect the green line, indicating that there is currently no evidence to suggest that these forecast days are more skillful than a daily 180-day prediction period at identifying the correct maximum daily GMS level.

Figure 3 displays Relative Operating Characteristic (ROC) plots calculated using GMSFs issued between April 2015 and October 2016; these plots describe the skill associated with each day of the forecast at correctly discriminating the days on which Kp reached or exceeded 5- (G1). A ROC curve is simply a plot of the Hit Rate (the proportion of events that were forecast) against the False Alarm Rate (the proportion of non-events that were incorrectly forecast), both of which range between 0 and 1. The Hit Rate is positively orientated whereas the False Alarm Rate is negatively orientated. Each point on a ROC curve represents the value of these two statistics at a different probability level. If action is taken when the forecast probability of an event is low the Hit Rate will be relatively large because events are forecast more frequently; however, the False Alarm Rate will also be relatively large because many of these forecasts will be false alarms. As the action/no-action forecast probability threshold is increased the value of both statistics reduce (tending to zero when action is never taken) and a ROC curve is formed by drawing a line through all these points. The further this curve resides above and to the left of the leading diagonal the more skill the forecast has at correctly distinguishing events from non-events; however, a curve that resides close to the diagonal indicates that the forecast cannot distinguish events from non-events. Although the WVS assesses the performance of all levels identified in the GMSF, only G1/2 is considered in the present study because the performance statistics associated with more severe levels are insufficiently robust for detailed analysis due to their low base rates.

There are three ROC-curves in each sub-plot, the green curve represents standard (un-flexed) verification methodology, whereas the black lines apply flexing using low-hit and low-miss categories. All the points within each sub-plot of Figure 3 reside above the grey diagonal no-skill line, indicating that each day of the GMSF has skill at discriminating events of G1 or above. The black line formed by + points awards a hit to any warning during which the maximum Kp value is at least 4- but only registers a missed event when G1 is not forecast and the maximum Kp value is at least 5-. Whereas the black line formed by □ points also awards a miss when Kp is at least 4- when no warning is issued. The purpose of the exclusive-flexed curve (+ points) is to give the forecaster the benefit of the doubt when Kp values occur which are almost classified as a G1, whilst not penalizing near-G1 events; whereas the inclusive-flexed curve (□ points) assesses whether the forecasts are inadvertently using a different threshold to Kp=5-. Comparing each pair of + and □ points at every probability threshold reveals that the Hit Rate values significantly increases whereas the False Alarm Rate remains virtually unchanged – this is a consequence of the low base rate. The curve formed by the + points indicates that during a significant proportion of the days on which G1 was predicted the maximum Kp value was either 4-, 4o or 4+.

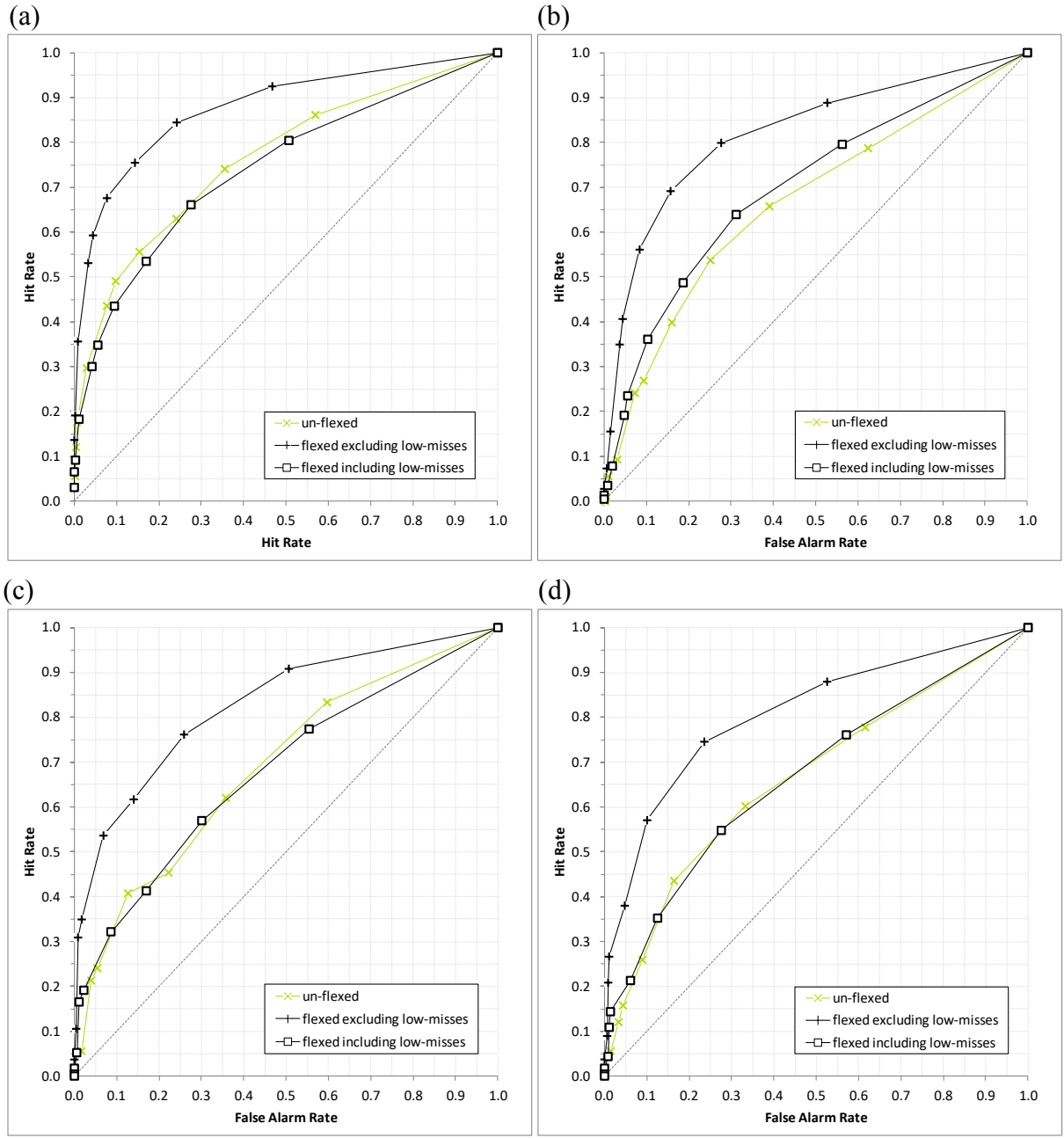

**Figure 3**. ROC-plots generated for GMS level events ≥ G1 using the (×) un-flexed, the flexed including low-misses (□) and the flexed excluding low-misses (+) technique for GMSFs on (a) day 1; (b) day 2; (c) day 3 and (d) day 4, issued between April 2015 and October 2016.

In each plot the exclusive-flexed curve (+) shows better discrimination than the green un-flexed curve. This clearly indicates that maximum daily Kp values of 4-, 4 or 4+ often occur when G1 is forecast with a non-zero probability. The inclusive-flexed curve (□) amounts to simply reducing the Kp event threshold to 4- (from 5-), it is interesting to observe that the resulting ROC-curves virtually coincide with the green un-flexed curves. A comparison of each point in sub-figure (a) reveals that inclusive-flexed values of the Hit Rate and False Alarm Rate are smaller than their green un-flexed counterparts. The fact that the curves are virtually coincident is an indication that the decrease in the proportion of correctly warned-for events is matched by an increase in the proportion of forecasts that were false alarms; consequently, the ability with which events are correctly identified is almost unchanged.  In other words, the GMSF is equally skilled at identifying days on which Kp≥4- as it is at identifying days on which Kp≥5-. The same conclusion also applies to sub-figures (c) and (d) (forecast days 3 and 4); however, sub-figure (b) appears to indicate that day 2 (identified as the worst performer in Figure 4) has (slightly) more skill at identifying Kp=4-.

The areas under the un-flexed (×), inclusive-flexed (□) and exclusive-flexed (+) ROC-curves on forecast day:

1. are 0.764, 0.742 and 0.878;
2. are 0.667, 0.699 and 0.822;
3. are 0.688, 0.680 and 0.827; and
4. are 0.664, 0.666 and 0.809 respectively.

Figure 4 displays reliability diagrams for the period between April 2015 and October 2016, to assess the accuracy of the probabilities with which the GMSF predicts days during which the maximum Kp value is at least 5- (G1). In both sub-figures: the grey dotted diagonal indicates perfect reliability; the region between the grey dashed lines denotes skill (in the Brier Score sense); the horizontal dot-dashed line of no-resolution denotes the frequency of occurrence of G1; and the solid, long-dashed, short-dashed and dotted back lines show the reliability of forecast days 1 through 4 respectively. The dark grey, mid grey, light grey and pale grey histograms denote the frequency with which events were predicted on days 1 through 4 respectively. In sub-figure (a) the verifying Kp event threshold is 5-, whereas in (b) it is 4-; consequently, these plots are the counterpart to the un-flexed (×) and inclusive-flexed (□) ROC-curves in Figure 3. The horizontal dot-dashed lines reveal that the maximum daily Kp value was ≥ 5- on 18% of occasions and ≥ 4- on 39% of occasions. The histograms (which are identical in both figures because the forecast is identical) reveal that G1 was rarely forecast with a high probability, especially at longer range. On days 1, 2, 3 and 4 probabilities ≥ 50% were issued on 17%, 13%, 9% and 7% of occasions and probabilities ≥ 90% on only 15, 3, 1 and 1 occasions respectively. Consequently, there is low confidence associated with the points in these figures that represent forecasts of higher confidence; never-the-less (with the exception of the 0-10% probability bin) almost all remaining points in sub-figure (a) lie below the no-skill region – a clear indication that G1 was over-predicted. The equivalent lines in sub-figure (b) lie above the dotted-diagonal (perfect skill) line because the event threshold used in this plot is a Kp-value of 4- (rather than 5-); however, although the majority of these points indicate under-forecasting many of them lie in the region between the two grey off-diagonal dashed lines (the forecast-skill-region). This appears to indicate that the forecast is more reliable at correctly identifying Kp values ≥ 4- than those ≥ 5-. The Brier score can be decomposed into three components (*Jolliffe and Stephenson, 2012*), one of these is a negatively orientated measure (between 0 and 1) of the reliability (*REL*) given by

$$REL = \sum_{k=1}^{K} \frac{n_k}{N} \left( \frac{o_k}{n_k} - P_k \right)^2. \qquad (5)$$

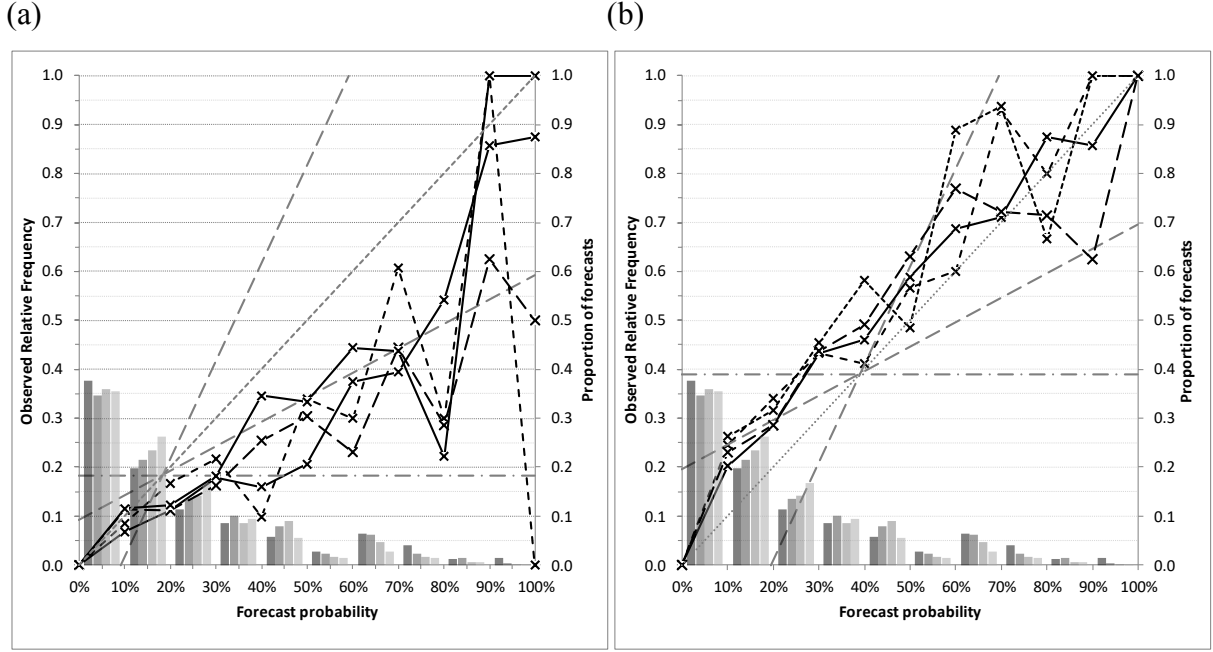

**Figure 4**. Reliability diagrams for GMSFs of G1 issued between April 2015 and October 2016 on: day 1 (solid/dark grey), day 2 (long-dashed/mid-dark grey), day 3 (short-dashed/mid-grey) and day 4 (dotted/light grey); when verified against daily maximum Kp values of at least (a) 5- and (b) 4-.

In this expression $K$ denotes a probability bin, $N$ is the total number of forecast days, $n_k$ is the number of times a geo-magnetic storm was forecast with a probability $P_k$ and $o_k$ is the total number of times a geo-magnetic storm was observed given that it was forecast with a probability $P_k$. REL for forecast day:

1. is 0.024 in (a) and 0.009 in (b);
2. is 0.025 in (a) and 0.014 in (b);
3. is 0.019 in (a) and 0.018 in (b);
4. is 0.013 in (a) and 0.023 in (b).

These scores (being negatively orientated) confirm that for forecast days 1 and 2 the GMSF appears to provide a slightly more reliable forecast for lower Kp (4- to 4+) events.

## 3.2 X-ray Flares

A similar analysis to that described in Section 3.1 was also undertaken to evaluate the most suitable rolling prediction period for the evaluation of a reference for XRF forecasts. Rolling mean RPS values were again used to identify 120 days as the best prediction period.

Examination of flare occurrence over solar cycle (see e.g., the histograms of Figure 5 in *Wheatland*, 2005) confirms that a relatively short prediction period is a sensible choice, since periods longer than 12 months would prove problematic during the sharp rising and declining phases; therefore, daily prediction period lengths between 30 and 360 days were considered. When undertaking this analysis for GMSFs the truth data were only available following an (approx) one-month latency period; however, the truth data source for XRFFs is GOES long-wave radiation flux and minute-by-minute values for these are available instantly. Therefore, no such latency period is appropriate because the truth data are immediately available to the forecaster. As was the case or GMSFs, extensive checking of every considered prediction period length appeared to produce similar 12-month rolling mean RPS values; however, in the present case, an examination of minimum mean RPS values on the first day of each month throughout the 16-year trial period did not reveal any optimal prediction period length. Therefore, again taking into consideration solar cycle variation as highlighted in Figure 5 of *Wheatland*, 2005 a 120-day prediction period was chosen.

Figure 5 displays the rolling 120 day frequency of occurrence of ABC-class (×), M-class (×) and X-class (□) flares from January 2001 and including the period of MOSWOC forecasts analyzed in the present study; it clearly reveals the extent to which ABC-class flares dominate and their frequency of occurrence has noticeably increased in recent months. During the past five years the minimum 120 day frequency of occurrence of ABC-class flares was 56.8% - observed in March 2014 and the maximum of 99.3% was observed in July 2016. During the period of SWTF analysis the maximum XRF class to occur in the 120 day prediction period was:

- ABC on between 70.0% and 99.2% of days;
- M on between 0.8% and 28.3% of days;
- X on between 0% and 1.7% of days.

Rolling 12-monthly RPSSs for each day of the XRFF (together with 90% confidence intervals, calculated using bootstrapping with replacement) are displayed in Figure 6. These scores have been evaluated via Equation (2), using the PDFs in Figure 5 to calculate $\overline{RPS}_{ref}$; forecast days 1 to 4 are shown as solid, long-dashed, short-dashed and dotted lines respectively. All point estimates of the RPSSs on days 1 and 2 lie above the green no-skill line, as do the majority of estimates on days 3 and 4; however, all their accompanying confidence intervals cross this line. Therefore, there is little evidence to suggest that the skill of the forecast at correctly identifying the maximum daily XRF class exceeds that obtained by using a rolling 120-day prediction period. Similarly the confidence intervals provide little evidence to suggest that any one forecast day is more skilful than another; however, the estimates alone suggest that day 1 tends to be more accurate than subsequent forecast days. What is obvious from this figure is the increasing size of the confidence intervals – those associated with the RPSS calculated for the 12-month period to March 2016 are much smaller than the equivalent values calculated in October 2016. One probable cause for this increase is the rapid decrease in the base rate of M-class flares over this time frame – as indicated by the green line in Figure 5.

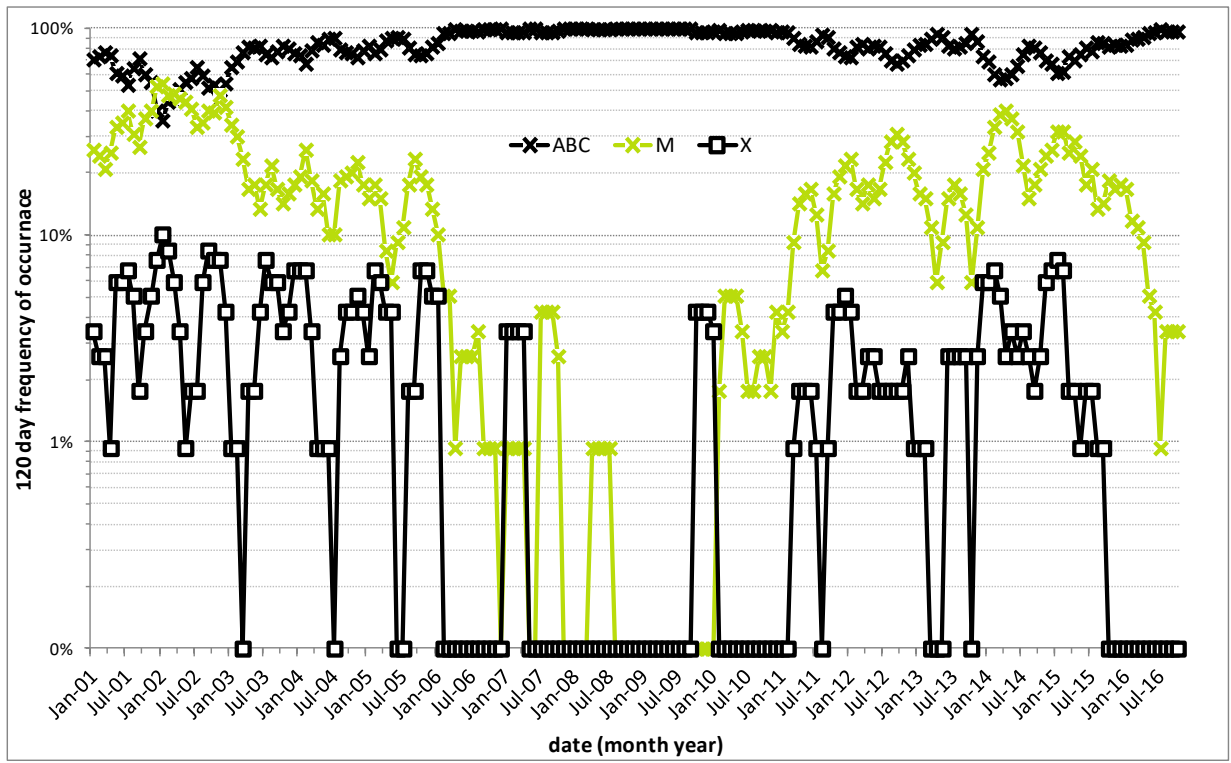

**Figure 5.** Rolling 120-day frequency of occurrence of daily maximum long wave radiative flux class.

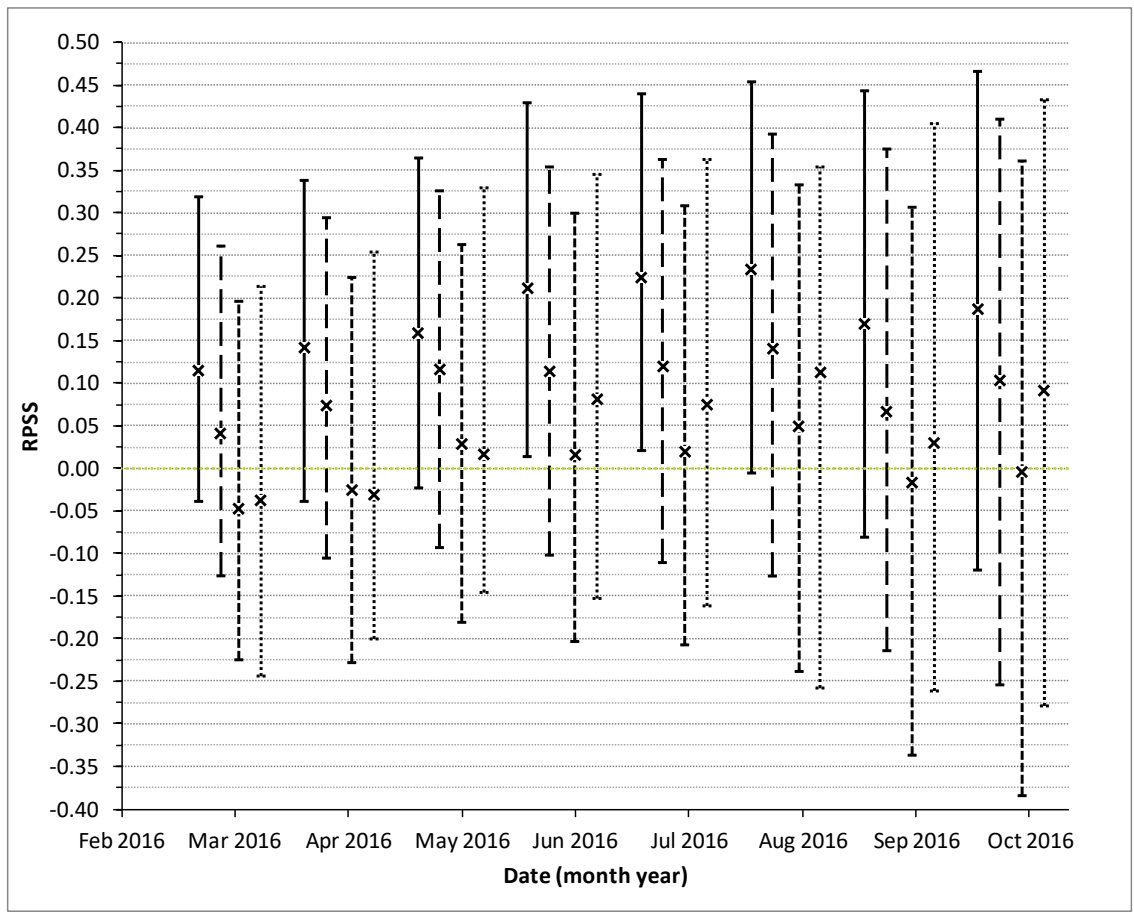

**Figure 6.** Rolling 12-monthly RPSS values with 90% bootstrapped confidence intervals for each day of the XRFF

Figure 7 displays ROC-plots calculated from XRFFs issued between April 2015 and October 2016; these plots describe the skill associated with each forecast day at correctly discriminating when the maximum daily flux is at least M-class. Although the WVS assesses the performance of both M and X class flares, only M is considered in the present study because the performance statistics associated with X are insufficiently robust for detailed analysis, due to their low base rate.

The three curves that are displayed in each sub-figure are as described in relation to Figure 3, except that in the present case the low-hit threshold corresponds to a C-class flare. The points on each curve of each sub-figure lie above the grey diagonal no-skill line, indicating that each day of the XRFF has skill at discriminating fluxes corresponding to M-class (or C-class) flares or above. In each plot the exclusive-flexed curve (+) displays better discrimination than the un-flexed curve (×), clearly indicating that a C-class flare often occurs when an M-class is forecast with a non-zero probability. The inclusive-flexed curve (□) amounts to simply reducing the event threshold to a C-class flare, and the resulting curves indicate less discriminatory skill compared with the un-flexed curves. It is likely that a reduction in the event threshold will increase: the base rate, the number of hits and the number of missed events; it is also likely to decrease the number of false alarms and correct rejections. In Figure 7 the Hit Rates on each inclusive-flexed curve are smaller than their un-flexed counterparts, indicating that as a proportion, the number of missed events has increased more than the number of hits. Inclusive-flexed False Alarm Rates have also reduced compared with their un-flexed counterparts, indicating that (as a proportion) the number of false alarms has reduced more than the number of correct rejections; however, this decrease is not large enough to offset the reduction in Hit Rate and consequently the area underneath the inclusive-flexed ROC-curve (a summary indicator of discriminatory skill) has reduced. The areas under the un-flexed, inclusive-flexed and exclusive-flexed ROC-curves on day:

1. are 0.874, 0.784 and 0.989;
2. are 0.871, 0.786 and 0.989;
3. are 0.847, 0.777 and 0.984; and
4. are 0.822, 0.775 and 0.980 respectively.

For each type of flexing the area under the ROC-curve decreases monotonically with increasing forecast range. This trend was also found in the *Murray et al* (2017) work, with the day 1 forecast generally being more skillful than subsequent days. The fact that the area beneath each inclusive-flexed ROC-curve is smaller than the corresponding area beneath each un-flexed ROC-curve indicates that either the C-class flare threshold ($1.0E^{-6}Wm^{-2}$) provides a low-hit threshold that is too small or that the discriminatory skill of the XRFF service is optimized by the M-class flare threshold ($1.0E^{-5}Wm^{-2}$). + points only register missed events when an M-class flare is not forecast and an M-class flare occurs, whereas a hit is awarded to any forecast during which the maximum XRF class is at least C; the purpose being to give the forecaster the benefit of the doubt when XRFs occur which are almost classified as M-class, whilst not penalizing flare events that were nearly M-class. Comparing each pair of + and □ points at every probability threshold reveals that the change in Hit Rates is significantly greater than the change in False Alarm Rates – this is a consequence of the low base rate associated with XRFs.

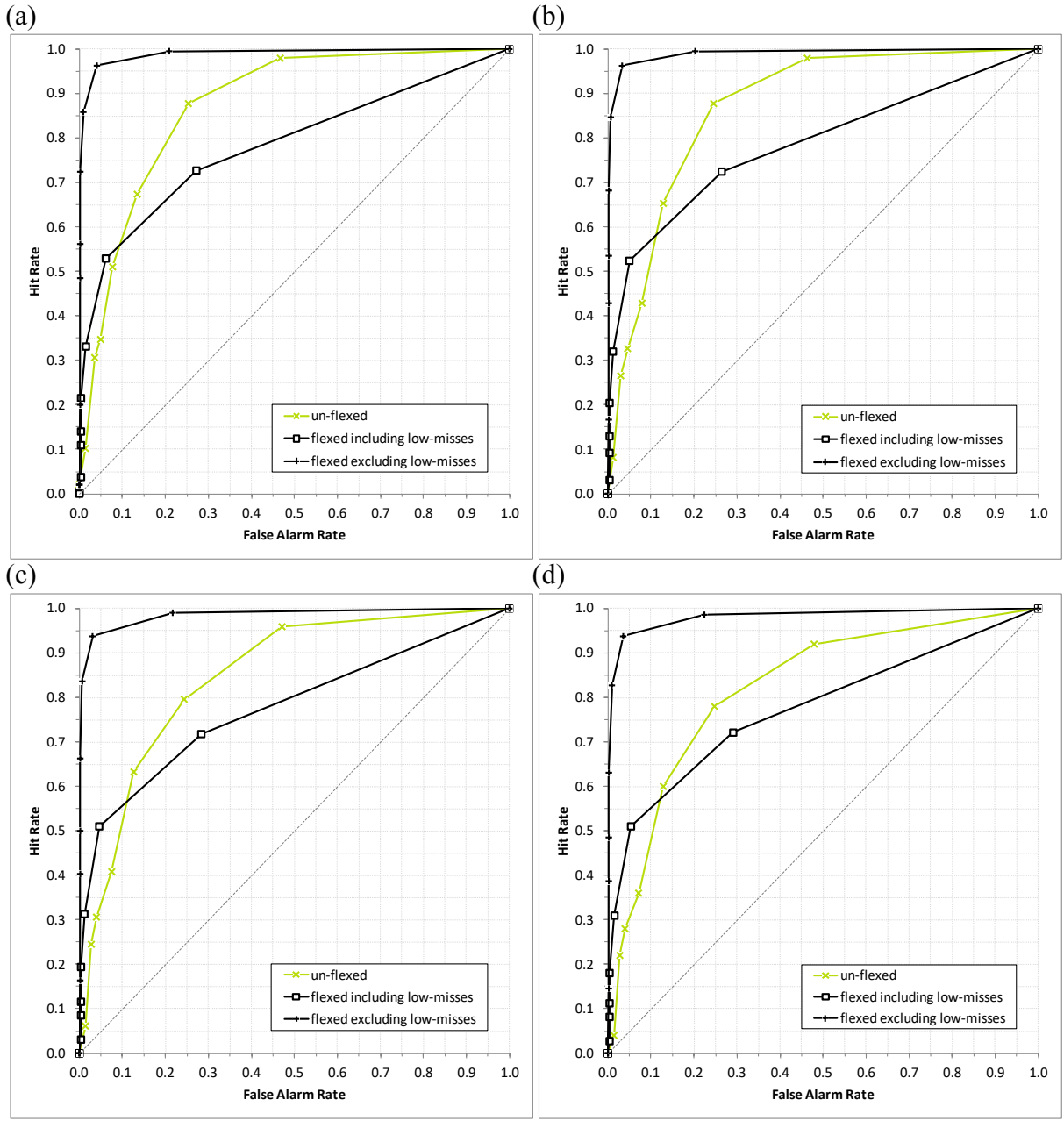

**Figure 7.** ROC-plots generated using the (×) un-flexed, the flexed including low-misses (□) and the flexed excluding low-misses (+) technique for (a) day 1; (b) day 2; (c) day 3 and (d) day 4 XRFFs for M-class flares, issued between April 2015 and October 2016.

Figure 8 displays reliability diagrams for the period between April 2015 and October 2016, these are used to assess the accuracy with which the XRFF predicted (a) M-class and (b) C-class flares: the format of these plots is identical to Figure 4. The horizontal dot-dashed lines indicate that XRFs of at least M-class and C-class occurred on 9% and 53% of occasions respectively.

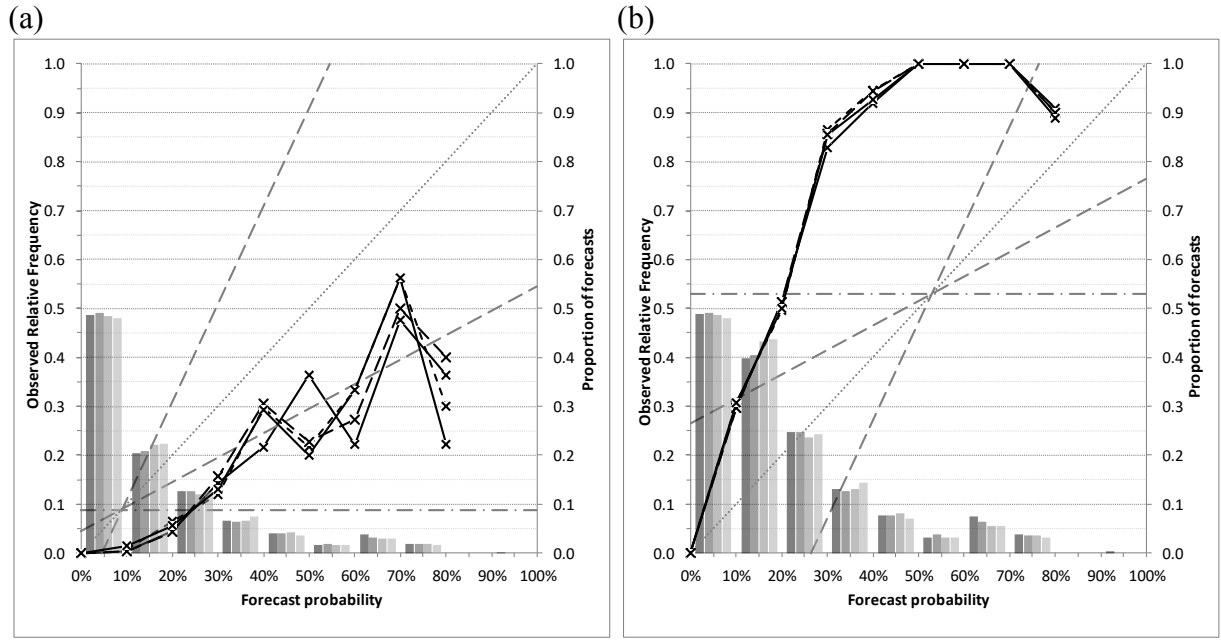

**Figure 8.** Reliability diagrams for XRFFs issued between April 2015 and October 2016 on: day 1 (solid/dark grey), day 2 (long-dashed/mid-dark grey), day 3 (short-dashed/mid-grey) and day 4 (dotted/light grey); when verified against daily maximum long-wave fluxes of at least (a) $1.0E^{-5} Wm^{-2}$ (M-class) and (b) $1.0E^{-6} Wm^{-2}$ (C-class).

The histograms reveal that M-class was rarely forecast with high probability; however, there is very little difference between the different shaded bars, indicating that the frequency with which M-class flares are forecast with n% probability (where n is a decile) is similar on each day of the forecast. In sub-figure (a) the majority of points lie below the no-skill region – a clear indication that M-class flares are over-forecast, as also found in the *Murray et al* (2017) study. However, the curve in sub-figure (b) is significantly above the diagonal indicating under-forecasting of C-class flares. Therefore, it appears that instead of using a long-wave flux threshold of $1.0E^{-5} Wm^{-2}$ (M-class) the actual (unintentional) threshold was between $1.0E^{-6} Wm^{-2}$ and $1.0E^{-5} Wm^{-2}$. The reliability component to the Brier Score for forecast day:

1. is 0.036 in (a) and 0.031 in (b);
2. is 0.032 in (a) and 0.031 in (b);
3. is 0.028 in (a) and 0.030 in (b);
4. is 0.028 in (a) and 0.029 in(b).

These negatively orientated scores are very similar; they appear to indicate that the XRFF predicts M-class and C-class flares with similar reliability.

## 4 Conclusions

The present study contains the results of analyzing GMSFs and XRFFs contained within daily 00Z SWTFs issued by MOSWOC over the 19-month period between April 2015 and October 2016. Two approaches have been adopted:

1. a ROC and reliability analysis is used to assess the ability with which G1 GMSs and M-class XRFs are predicted; and
2. a RPSS analysis is performed to analyse the skill of the GMSFs and XRFFs against the skill demonstrated by simply forecasting the frequency of occurrence over the most recent 180-day and 120-day prediction periods respectively (chosen to optimise $\overline{RPS_{ref}}$).

For the GMSF:

- The ROC analysis revealed that each day of the forecast had skill at discriminating days on which the maximum Kp-value was greater than or equal to 5- (G1); however, the forecast displayed a virtually identical level of skill at identifying days on which the maximum Kp-value was greater than or equal to 4-.
- The reliability analysis revealed that G1 storms were over-forecast, whereas Kp-values ≥5- were slightly under-forecast; consequently, the GMSF was found to more reliably predict maximum Kp-values ≥4- than maximum Kp-values ≥5-.
- The RPSS analysis presented little statistically significant evidence that day 1 of the GMSF was a better predictor of maximum GMS level than the frequency of occurrence over the preceding 180 days.

For the XRFF:

- The ROC analysis revealed that each day of the forecast had more skill at correctly identifying M-class flares than C-class flares.
- The reliability analysis confirmed that although M-class flares are over-forecast, C-class flares are greatly under-forecast; therefore, it is likely that the most appropriate event-threshold was between $1.0E^{-6}Wm^{-2}$ (C-class) and $1.0E^{-5}Wm^{-2}$ (M-class).
- The RPSS analysis indicated that the XRFF struggled to outperform a forecast comprised of only the frequency of occurrence over the preceding 120 days, with the confidence intervals associated with these estimates providing no statistically significant evidence.

In the future our goals are to continue the analysis of the GMSF and XRFF components to the SWTF as this provides valuable feedback and guidance to MOSWOC forecasters. Plans also exist to: compare the performance of these services against equivalent services provided by other space weather centres and expand the verification to include other SWTF components (the next area of study being coronal mass ejection forecasts).

## Acknowledgments, Samples, and Data

The authors would like to thank the MOSWOC forecasting team and David Jackson and Suzy Bingham from the Space Weather Research Team for their assistance. S. A. Murray was partly supported by the European Union Horizon 2020 research and innovation programme under grant agreement No. 640216 (FLARECAST project).

All data necessary to reproduce the findings in this study are freely available for:

- observed planetary Kp values via FTP download GFZ Helmholtz centre website http://www.gfz-potsdam.de/en/home/

- observed long wave radiation flux reported from GOES via FTP download from the SWPC website http://www.swpc.noaa.gov/
- MOSWOC Space Weather Technical Forecasts via the Met Office http://www.metoffice.gov.uk/ using a Freedom of Information request

## Glossary of Terms

AFVS: Area Forecast Verification System
CME: Coronal Mass Ejection
FLARECAST: Flare Likelihood And Region Eruption foreCASTing
GOES: Geo-Orbiting Earth Satellite
GMSF: Geomagnetic storm forecasts
MOSWOC: Met Office Space Weather Operations Centre
PDF: Probability Density Function
ROC: Relative Operating Characteristic
RPS: Ranked Probability Score
RPSS: Ranked Probability Skill Score
SWPC: Space Weather Prediction Centre
SWTF: Space Weather Technical Forecast
XRFF: X-Ray Flare Forecasts
WVS: Warnings Verification System